\documentclass[11pt]{article}
\usepackage{epsf}
\usepackage{axodraw}

\newcommand{ \centeron }[2]{{\setbox0=\hbox{#1}\setbox1=\hbox{#2}\ifdim
                             \wd1>\wd0\kern.5\wd1\kern-.5\wd0\fi \copy0
                             \kern-.5\wd0\kern-.5\wd1\copy1\ifdim\wd0>\wd1
                             \kern.5\wd0\kern-.5\wd1\fi}}
\newcommand{ \ltap }{\>\centeron{\raise.35ex\hbox{$<$}}
                     {\lower.65ex\hbox{$\sim$}}\>}
\newcommand{ \gtap }{\>\centeron{\raise.35ex\hbox{$>$}}
                     {\lower.65ex\hbox{$\sim$}}\>}
\newcommand{ \gsim }{\mathrel{\gtap}}
\newcommand{ \lsim }{\mathrel{\ltap}}

\newcommand{ \slashchar }[1]{\setbox0=\hbox{$#1$}   
   \dimen0=\wd0                                     
   \setbox1=\hbox{/} \dimen1=\wd1                   
   \ifdim\dimen0>\dimen1                            
      \rlap{\hbox to \dimen0{\hfil/\hfil}}          
      #1                                            
   \else                                            
      \rlap{\hbox to \dimen1{\hfil$#1$\hfil}}       
      /                                             
   \fi}                                             %

\newcommand{ \Mpl      }{M_{\rm Pl}}

\newcommand{ \ra       }{\rightarrow}
\newcommand{ \textfrac }[2]{ {\textstyle\frac{#1}{#2}} }
\newcommand{ \ph       }{\gamma}

\newcommand{\drawsquare}[2]{\hbox{%
\rule{#2pt}{#1pt}\hskip-#2pt
\rule{#1pt}{#2pt}\hskip-#1pt
\rule[#1pt]{#1pt}{#2pt}}\rule[#1pt]{#2pt}{#2pt}\hskip-#2pt
\rule{#2pt}{#1pt}}
\newcommand{\Yfund}{\raisebox{-.5pt}{\drawsquare{6.5}{0.4}}}

\def\singleandabitspaced{\baselineskip=\normalbaselineskip\multiply
    \baselineskip by 110\divide\baselineskip by 100}
\def\abstractspacing{\baselineskip=\normalbaselineskip\multiply
    \baselineskip by 110\divide\baselineskip by 100}
\def\singlespaced{\baselineskip=\normalbaselineskip}

\begin{document}

\singlespaced

\begin{titlepage}

\begin{flushright}
hep-ph/0104074 \\
MADPH--01--1219 \\
\end{flushright}

\vspace{1.5cm}

\begin{center}
\mbox{\LARGE \textbf{Radion effects on unitarity in 
gauge-boson scattering}} \\

\vspace*{2.0cm}
{\Large Tao Han, Graham D. Kribs, and Bob McElrath} \\
\vspace*{0.5cm}
\textit{Department of Physics, University of Wisconsin, \\
        1150 University Ave., Madison, WI~~53706-1390} \\

\vspace*{1.0cm}

\texttt{than@pheno.physics.wisc.edu, kribs@pheno.physics.wisc.edu,
mcelrath@pheno.physics.wisc.edu}

\vspace*{1.0cm}

\begin{abstract}
\indent

\abstractspacing

The scalar field associated with fluctuations in the positions 
of the two branes, the ``radion'', plays an important role determining 
the cosmology and collider phenomenology of the Randall-Sundrum solution 
to the hierarchy problem.  It is now well known that the radion mass is 
of order the weak scale, and that its couplings to standard model 
fields are order ${\cal O}({\rm TeV}^{-1})$ to the trace of the energy 
momentum tensor.  We calculate longitudinal vector boson scattering 
amplitudes to explore the constraints on the radion mass and its
coupling from perturbative unitarity.  The scattering cross section 
can indeed become non-perturbative at energies prior to reaching the 
TeV brane cutoff scale, but only when some curvature-Higgs mixing 
on the TeV brane is present.  We show that the coefficient of the 
curvature-Higgs mixing operator must be less than about $3$ for 
the 4-d effective theory to respect perturbative unitarity up to 
the TeV brane cutoff scale. Mass bounds on the Higgs boson
and the radion are also discussed.

\end{abstract}

\end{center}
\end{titlepage}

\newpage
\setcounter{page}{2}
\renewcommand{\thefootnote}{\arabic{footnote}}
\setcounter{footnote}{0}
\singleandabitspaced

\section{Introduction}
\label{introduction-sec}

Theories with extra dimensions have received tremendous attention
within the last a few years.  One of the most interesting 
incarnations was formulated by Randall and Sundrum (RS1) \cite{RS1}, 
who postulated a universe with two 4-d surfaces (``branes'') bounding a 
slice of 5-d AdS spacetime.  The Standard Model (SM) fields are assumed
to be located on one brane (the ``TeV brane'').  Gravity lives on the 
other brane (the ``Planck brane'') and in the bulk as well as the TeV brane.  
Both branes have equal but opposite tension, 
while the bulk has a (negative) cosmological constant.  By carefully
tuning the brane tensions against the bulk cosmological constant,
one can achieve a low energy effective theory that has flat 4-d spacetime.
The RS1 metric takes the form\footnote{In our
conventions, the metric has signature ($+$,$-$,$-$,$-$), Greek 
indices $\mu,\nu,$ etc., run from $0,1,2,3$ denoting ordinary 4-d 
spacetime.  The extra dimension is assumed to be compactified on 
a $S_1/Z_2$ orbifold.}
\begin{eqnarray}
d s^2 &=& e^{-2 k L y} \eta_{\mu \nu} dx^{\mu} dx^{\nu} - L^2 dy^2
\end{eqnarray}
where $L$ is the size of the extra dimension and $0 \le y \le 1$.
All mass scales in the full 5-d theory are of order the Planck scale.
By placing the SM fields at $y=1$, all mass terms must be rescaled
by an exponential suppression factor (``warp factor'') $e^{-k L}$ that 
can bring them down to the TeV scale.  This merely requires that 
$L \sim 35/k$, and thus roughly $35$ times the fundamental Planck length.
This is a dramatic improvement over the original hierarchy problem
between the electroweak scale and the 4-d Planck scale $\Mpl$.

The first obvious difficulty with this scenario is to arrange
that the extra dimension stabilizes to a size of about an order
of magnitude larger than the Planck length.  In the original
proposal, the potential for the size of the extra dimension 
is flat, so that all sizes are classically equivalent.
The actual size of the extra dimension was tuned 
appropriately to solve the hierarchy problem. 
A more serious concern was first identified by Ref.~\cite{Csaba1}, 
in which enforcing $d L(t)/dt = 0$ 
in a cosmological context implied a nontrivial relationship
between the TeV and Planck brane energy densities.  
Ultimately this was shown to be a direct result of assuming the 
potential for the size of the extra dimension is flat.  Hence, the
scalar field associated with fluctuations of the size of the
extra dimension, the ``radion'', is massless.  If, on the other 
hand, \emph{bulk} dynamics setup a potential whose minimum 
determined the distance separating the two branes, then the
radion acquires a mass. 

Goldberger and Wise (GW) \cite{GWstable} noticed that a potential 
could be setup for the radion, by adding a 5-d bulk scalar to RS1 
arranged so that it acquires an $y$-dependent vacuum expectation value (vev).
Furthermore, the classical potential is stable against quantum 
corrections \cite{GoldRoth}.
In their analysis, the 5-d metric was generalized to
\begin{eqnarray}
d s^2 &=& e^{-2 k L(x) y} \eta_{\mu \nu} dx^{\mu} dx^{\nu} - L(x)^2 dy^2 \; ,
\label{naive-eq}
\end{eqnarray}
where $L$ has a vev and an $x_{\mu}$-dependent fluctuation.
This generalization with radial fluctuations, however, does not 
satisfy Einstein's equations, and so a different ansatz for
the metric is needed to calculate the mass and couplings of the radion.
Charmousis, Gregory, and Rubakov \cite{CharmGregRub} proposed the metric 
\begin{eqnarray}
d s^2 &=& e^{-2 A(y) - 2 F(x)} \eta_{\mu\nu} dx^{\mu} dx^{\nu}
 - \left( 1 + 2 F(x) \right) dy^2
\end{eqnarray}
in which they showed that a consistent treatment of the
radial fluctuations characterized by the scalar field $F(x)$
is possible while solving the linearized Einstein equations.  
(In the Randall-Sundrum model, $A(y) = k L y$.)

This analysis, however, did not take into
account the effect of the bulk scalar field back onto the metric
(the ``back-reaction''). 
Including the back-reaction of the bulk scalar field on the
metric is in general highly nontrivial.  However, DeWolfe, Freedman, 
Gubser, and Karch \cite{DeWolfe} proposed an interesting
generating solution technique motivated from gauged supergravity.  
Their technique allows one to calculate the potential for the
size of the extra dimension consistently including the effects
of the bulk scalar field vev profile into the metric.  This
manifested itself through additional $y$-dependent terms in $A(y)$.
However, radial fluctuations were not considered.

This motivated the work by Cs\'aki, Graesser, and Kribs \cite{CGK}
(see also \cite{TanakaMontes})
that used the generating solution technique by
DeWolfe et al.\ \cite{DeWolfe}, combined with the CGR metric
ansatz \cite{CharmGregRub}, to show that a consistent treatment of 
the fluctuations in the size of the extra dimension could be done.
In this work, the wave-function, mass, and couplings of the radion
were explicitly calculated.  In particular, it was shown that 
the mass of radion is of order TeV/35, multiplied by the size
of the back-reaction (taken as a perturbation).  Remarkably, this closely
matched the radion mass calculation in Refs.~\cite{GWstable, Csaba2, 
GWpheno}, that used the ``naive'' ansatz in Eq.~(\ref{naive-eq}).  
The radion couplings could be obtained through 
\begin{eqnarray}
r \frac{\delta S}{\delta r} = 
   r \frac{\delta S}{\delta g_{\mu\nu}} \frac{\delta g_{\mu\nu}}{\delta r} = 
   \frac{r}{e^{-kL} \Mpl} T_\mu^\mu
\end{eqnarray}
and thus couples
to the trace of the energy momentum tensor $T_{\mu\nu}$.  Here $r$ is the
canonically normalized radion field, related to $F(x)$ via \cite{CGK}
\begin{eqnarray}
F(x) &=& \frac{1}{\sqrt{6} \Mpl e^{-k L}} r(x) \; .
\end{eqnarray}
The strength of the radion couplings are proportional to 
the inverse ``warped'' Planck scale, and thus 
for appropriate choice of $L$, the TeV scale.  One of the new results 
of this analysis is that the Kaluza-Klein (KK) excitations of the 
bulk scalar field couple to the SM fields, albeit suppressed by the 
size of the back-reaction divided by the KK mass (of order the warped
Planck scale).

With a radion mass that is
of order the weak scale, and couplings that are of order
1/TeV, several groups proceeded to analyze the phenomenology of 
the radion \cite{GRW2,BaeFeb2000,radionpheno}.  
It was quickly realized that
the radion couplings are analogous to the Higgs at tree-level.
(At one loop, the radion couples to the trace anomaly \cite{GRW2,BaeFeb2000}, 
and therefore has a significantly different coupling to, for example, 
massless gauge bosons.) 
In particular, the tree-level couplings of the radion to 
the electroweak gauge bosons are the same as those of the SM Higgs, 
upon substituting 
\begin{eqnarray}
h \longrightarrow -\gamma r, \; \ {\rm with}\ \ 
\gamma = \frac{v}{\sqrt{6} \Lambda} \; ,
\label{hr}
\end{eqnarray}
where $h,r$ are the Higgs and radion mass eigenstates,
$v = 246$ GeV is the electroweak vev, and the warped Planck scale
is $\Lambda = e^{-k L} \Mpl$ \cite{Csaba2,GWpheno,CGK}.  
Hereafter, we will use ``TeV brane cutoff scale'' or just
``cutoff scale'' to refer to the warped Planck scale,
since our 4-d effectively theory is valid only up to about $\Lambda$.
This leads to new contributions to the electroweak 
precision observables, such as the oblique corrections \cite{CGK}.
Generally, the size of this effect is rather small, since 
for a cutoff scale of order a TeV, $\gamma$ is order $0.1$. 
For much lower cutoff scales, direct KK graviton production
is important, and can provide constraints on the RS1
scenario \cite{DHR}.

Since the radion couplings to the SM are similar to those of
the Higgs boson, it is natural to ask if the radion has any
significant effects on the electroweak symmetry breaking sector,
especially in the pessimistic scenario where the SM Higgs
is rather heavy and may not be easily produced at collider
experiments.
In this paper we consider the effects of the radion on perturbative 
unitarity bounds in the SM\@.  A few papers 
\cite{Mahantaunitarity,Baeunitarity}
have considered some of the effects of the radion on unitarity
involving external Higgs bosons, although no explicit description 
of the Goldstone boson
equivalence theorem is present, nor the effects of including
curvature-Higgs mixing.  The paper is organized as follows.  
In Sec.~\ref{unitarity-sec} we give a brief discussion of 
unitarity issues in the SM, and the bound on the Higgs mass
that results.  In Sec.~\ref{radioneffects-sec} we introduce 
the 4-d effective theory that includes the radion and write
the relevant interactions for gauge boson scattering.
In Sec.~\ref{W-scat-sec} we calculate the partial wave amplitude
including the effects of the radion to the largest process in the SM, 
namely $W^+W^- \rightarrow W^+W^-$, and show that there is in general 
no significant constraint on the radion mass or coupling in the absence 
of other interactions.  In Sec.~\ref{radion-gold-sec} we explicitly
demonstrate that the Goldstone boson equivalence theorem can
be applied, and thus in the high energy limit (large $s$) one obtains 
the same result replacing the longitudinally polarized $W$'s by the eaten
Goldstone bosons.  The above analysis, however, neglected curvature-Higgs
Higgs mixing (localized on the TeV brane).  In 
Sec.~\ref{curv-scalar-sec}, we introduce the mixing, 
and recalculate the partial wave amplitude.  We find that 
with a mixing coefficient $|\xi| \gsim 2.7$, the partial
wave amplitude for $W$ scattering does exceed the unitarity bound 
for scattering energies \emph{lower} than the cutoff scale.  
Finally, in Sec.~\ref{conclusions-sec} we present our conclusions.

\section{Perturbative unitarity}
\label{unitarity-sec}

In the SM, the longitudinal components of the electroweak 
gauge bosons ($W^\pm_L,Z_L$) arise from the eaten Goldstone bosons 
resulting from the spontaneous breaking of the electroweak gauge symmetry. 
The study of scattering of
longitudinally polarized gauge bosons would thus be the most
direct means to explore the mechanism of the electroweak
symmetry breaking.  In this section, we briefly review
the physics with longitudinal gauge boson scattering in
the SM and discuss perturbative unitarity bounds.  This serves as 
the basis for our further study including the radion.

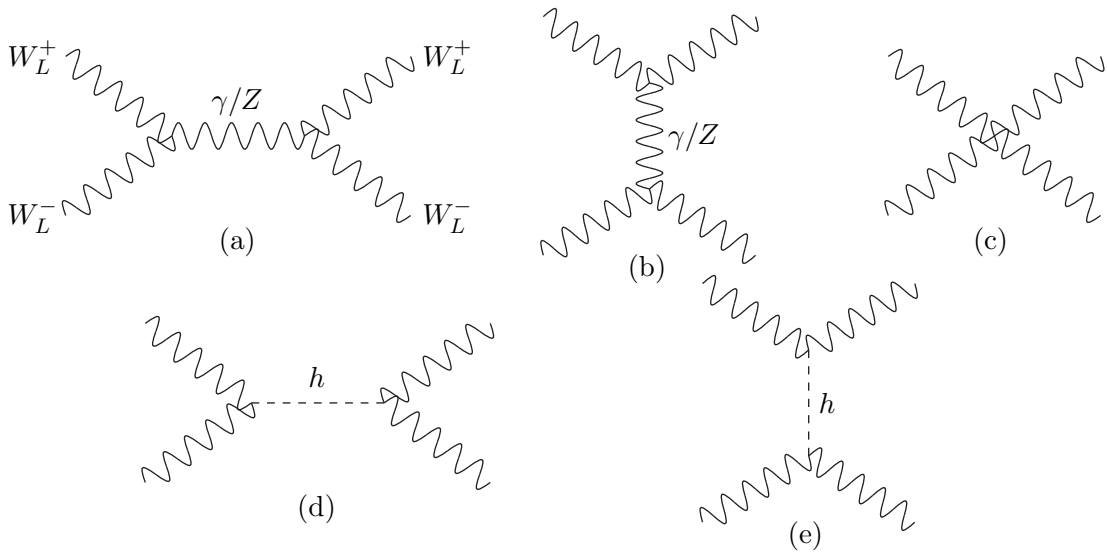
\begin{figure}[t]
\begin{picture}(440,100)
%
  \Text(    17, 80 )[r]{$W^+_L$}
  \Text(    17, 20 )[r]{$W^-_L$}
  \Photon(  20, 80 )(  60, 50 ){5}{5} 
  \Photon(  20, 20 )(  60, 50 ){5}{5} 
  \Photon(  60, 50 )( 110, 50 ){5}{5}
  \Photon( 110, 50 )( 150, 80 ){5}{5} 
  \Photon( 110, 50 )( 150, 20 ){5}{5} 
  \Text(    85, 63 )[c]{$\ph/Z$}
  \Text(   155, 80 )[l]{$W^+_L$}
  \Text(   155, 20 )[l]{$W^-_L$}
  \Text(    85, 10 )[c]{(a)}
%
  \Photon( 200, 95 )( 240, 70 ){5}{5}
  \Photon( 240, 70 )( 280, 95 ){5}{5}
  \Photon( 240, 70 )( 240, 30 ){5}{5}
  \Photon( 200, 05 )( 240, 30 ){5}{5}
  \Photon( 240, 30 )( 280, 05 ){5}{5}
  \Text(   248, 50 )[l]{$\ph/Z$}
  \Text(   240,  0 )[c]{(b)}
%
  \Photon( 330, 80 )( 370, 50 ){5}{5}
  \Photon( 370, 50 )( 410, 80 ){5}{5}
  \Photon( 330, 20 )( 370, 50 ){5}{5}
  \Photon( 370, 50 )( 410, 20 ){5}{5}
  \Text(   370, 10 )[c]{(c)}
\end{picture}
\begin{picture}(440,100)
%
  \Photon(    50, 80 )(  90, 50 ){5}{5} 
  \Photon(    50, 20 )(  90, 50 ){5}{5} 
  \DashLine(  90, 50 )( 140, 50 ){3}
  \Photon(   140, 50 )( 180, 80 ){5}{5} 
  \Photon(   140, 50 )( 180, 20 ){5}{5} 
  \Text(     115, 60 )[c]{$h$}
  \Text(     115, 10 )[c]{(d)}
%
  \Photon(   260, 95 )( 300, 70 ){5}{5}
  \Photon(   300, 70 )( 340, 95 ){5}{5}
  \DashLine( 300, 70 )( 300, 30 ){3}
  \Photon(   260, 05 )( 300, 30 ){5}{5}
  \Photon(   300, 30 )( 340, 05 ){5}{5}
  \Text(     305, 50 )[l]{$h$}
  \Text(     300,  0 )[c]{(e)}
\end{picture}
\caption{Tree-level Feynman diagrams for $WW$ scattering
in the SM.}
\label{W-scat-feyn-diag-fig}
\end{figure}

We focus on the process 
\begin{equation}
W^+_L W^-_L \rightarrow W^+_L W^-_L
\end{equation}
since, in the high energy limit, this gives the largest contribution 
to the partial wave amplitude of all $2 \rightarrow 2$ electroweak 
gauge boson scattering processes in the SM.
The Feynman diagrams are shown in 
Fig.~\ref{W-scat-feyn-diag-fig}.
At high energies, the wave-function of a longitudinal gauge
boson can be written as
\begin{eqnarray}
\epsilon^{W_L}_\mu(p) &=& \frac{p_\mu}{M_W} + 
{\cal O}\left( \frac{M_W}{E} \right) \; ,
\end{eqnarray}
This indicates that the longitudinally polarized gauge bosons are
behaving as their eaten Goldstone boson counterparts with corrections of
order $M_W/E$ where $E$ is the gauge boson energy.

With this substitution, the contributions from the diagrams in 
Fig.~\ref{W-scat-feyn-diag-fig}(a), (b), or (c) individually
leads to a contribution ${\cal O}(s^2/M_W^4)$, but these 
terms cancel among each other due to electroweak gauge 
symmetry.  The leading order result for the sum of these three diagrams,
neglecting ${\cal O}(M_W^2/s)$ terms, is
\begin{eqnarray}
-i {\cal M} &=& - \frac{u}{v^2} \; .
\label{gb-scat-eq}
\end{eqnarray}
At some point, the contribution to the partial 
wave amplitude is sufficiently large that the partial
wave unitarity will be violated.  This can be seen from the
zeroth order partial wave amplitude
\begin{eqnarray}
a_0 &=& \frac{1}{32\pi} \int_{-1}^1 d\cos\theta (-i{\cal M}) \; = \;
-\frac{s}{32\pi v^2} \; .
\label{part-eq}
\end{eqnarray}
If we interpret the partial wave unitarity bound to be
\begin{eqnarray}
\left|{\rm Re}(a_0)\right| &<& \frac{1}{2} \; ,
\end{eqnarray}
then one obtains the bound $\sqrt s < 1.7$ TeV.  This is
commonly referred as the indication of a strongly
interacting electroweak sector at the TeV scale \cite{cg}. 
On the other hand, by including the Higgs contribution 
of Figs.~\ref{W-scat-feyn-diag-fig}(d) and (e), the
amplitude becomes
\begin{eqnarray}
-i {\cal M}_{SM} &=& - \frac{m_h^2}{v^2} 
    \left( \frac{s}{s - m_h^2} + \frac{t}{t - m_h^2} \right) \; \to \; 
    - \frac{2 m_h^2}{v^2} \quad {\rm for} \quad m_h^2 \ll s,|t| \; .
\end{eqnarray}
A light Higgs boson thus naturally restores partial wave unitarity,
through a precise cancellation of the high energy divergence of the
gauge boson scattering amplitude against the additional physical Higgs 
contribution.  This mysterious cancellation becomes transparent
by simply replacing the full massive spin-1 
external gauge boson fields by the scalar Goldstone bosons,
as dictated by the equivalence theorem \cite{lqt,cg,et}.
However, this also implies an upper bound on the Higgs 
boson mass $m_h < 900$ GeV if perturbative unitarity is 
maintained to arbitrarily high energies \cite{lqt}.

Although the unitarity bound cannot predict any specific
form of new physics in the electroweak symmetry breaking
sector, it does provide a general argument for a scale
(around 1 TeV) at which the physics responsible for EWSB
must show up. Furthermore, any new physics that couples
to the EWSB sector significantly at high energies may be
subject to constraints from partial wave unitarity.

\section{The radion effects}
\label{radioneffects-sec}

In Randall-Sundrum scenarios, the low energy 4-d effective theory
includes the ordinary 4-d graviton, the radion, and the KK
excitations of the graviton and the bulk scalar.  Here we are assuming
the SM remains on the TeV brane (see e.g., Refs.~\cite{DHR,RS-SM-inbulk}
for alternatives).  At energies well below 
the cutoff scale, we need only consider the radion as the additional
degree of freedom in our effective theory.  As we approach (or exceed) 
the cutoff scale $\Lambda$, additional contributions from the KK modes 
(and perhaps other quantum gravity states) must be included in the 
calculations.  However, including these contributions is subtle and 
model-dependent.
For example, in the holographic viewpoint \cite{holographic}, the 
holographic dual 4-d theory is becoming strongly coupled in the 
transition to a 4-d conformal field theory above $\Lambda$.  
In any case, we will restrict ourselves to energies below the cutoff 
scale, and consider only the contribution from the radion.

As stated in the introduction, the radion couples to fields
localized on the SM brane through
\begin{eqnarray}
- \frac{\gamma}{v} \, r \, T_\mu^\mu \label{trace-eq}
\end{eqnarray}
which leads to the three-point interaction terms 
\begin{eqnarray}
{\cal L} &=& - \gamma \frac{2 M_W^2}{v} \, r \, W_\mu^+ W^{- \mu} 
             - \gamma \frac{M_Z^2}{v} \, r \, Z_\mu Z^\mu 
\label{rww-eq}
\end{eqnarray}
between the radion and the electroweak gauge bosons.
The Lagrangian written above is not complete \cite{CGK},
since there are additional gauge-fixing terms that must be
added.  In addition, in dimensional regularization the Lagrangian
is continued into $d$ dimensions, and there are yet more terms
with coefficients that depend explicitly on $(d-4)$.
Both of these sets of terms, however, will not be relevant here since
we are exclusively considering tree-level scattering amplitudes.
We now proceed to calculate the scattering amplitude using 
gauge bosons, and then compare by doing the same calculation using 
Goldstone bosons.

\subsection{Radion contributions to $W^+_L W^-_L$ scattering}
\label{W-scat-sec}

Since the radion couplings of Eq.~(\ref{rww-eq})
are analogous to the Higgs, there
are additional contributions to electroweak gauge boson 
scattering from radion exchange. The radion contributes
two additional diagrams 
to $W^+_L W^-_L \rightarrow W^+_L W^-_L$ scattering,
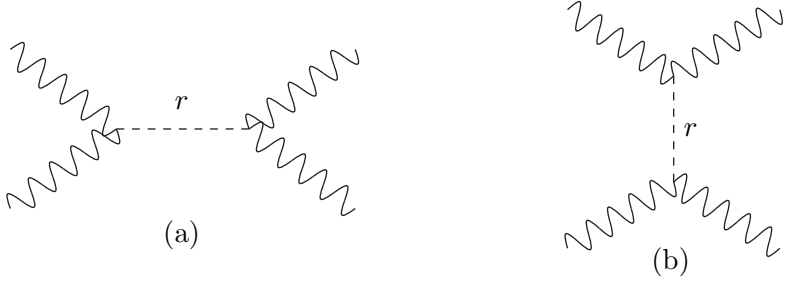
\begin{figure}[t]
\begin{picture}(440,100)
%
  \Photon(    50, 80 )(  90, 50 ){5}{5} 
  \Photon(    50, 20 )(  90, 50 ){5}{5} 
  \DashLine(  90, 50 )( 140, 50 ){3}
  \Photon(   140, 50 )( 180, 80 ){5}{5} 
  \Photon(   140, 50 )( 180, 20 ){5}{5} 
  \Text(     115, 60 )[c]{$r$}
  \Text(     115, 10 )[c]{(a)}
%
  \Photon(   260, 95 )( 300, 70 ){5}{5}
  \Photon(   300, 70 )( 340, 95 ){5}{5}
  \DashLine( 300, 70 )( 300, 30 ){3}
  \Photon(   260, 05 )( 300, 30 ){5}{5}
  \Photon(   300, 30 )( 340, 05 ){5}{5}
  \Text(     305, 50 )[l]{$r$}
  \Text(     300,  0 )[c]{(b)}
\end{picture}
\caption{Tree-level Feynman diagrams for $W$ scattering through
a radion.}
\label{radion-W-feyn-diag-fig}
\end{figure}
as shown in Fig.~\ref{radion-W-feyn-diag-fig}.
The amplitude for the sum of the two contributions at high energy,
neglecting ${\cal O}(M_W^2/s)$ terms as before, is
\begin{eqnarray}
-i {\cal M}_r &=& -g^2 \gamma^2 \Bigg[ \frac{s}{4 M_W^2} + \frac{t}{4 M_W^2}
    + \frac{m_r^2}{2 M_W^2} 
    + \frac{m_r^2}{4 M_W^2} 
    \left( \frac{m_r^2}{s - m_r^2} + \frac{m_r^2}{t - m_r^2} \right) 
\nonumber \\ & &{} \qquad \quad
    + 2 \frac{t}{s} 
    + \left( \frac{M_W^2}{m_r^2} + 1 - 2 \frac{m_r^2}{s} \right) 
    \frac{m_r^2}{s - m_r^2} 
    + \left( \frac{M_W^2}{m_r^2} + 1 + 2 \frac{m_r^2}{s} \right) 
    \frac{m_r^2}{t - m_r^2} \Bigg] \; .
\label{radion-amp-eq}
\end{eqnarray}
In the limit of large $s$, the leading order behavior of this amplitude
is $s/\Lambda^2$, irrespective of the radion mass.  This suggests that 
for sufficiently large $s$, the radion contribution is not perturbatively
calculable and thus may violate unitarity.
Following the analysis discussed in the previous section,
the zeroth partial wave is easily obtained:
\begin{eqnarray}
a_0 &=& -\frac{g^2}{16 \pi} \gamma^2 \Bigg[ \frac{s}{8 M_W^2} - \frac{3}{2}
    + \frac{m_r^2}{2 M_W^2}
    + \left( \frac{M_W^2}{m_r^2} + 1 + \frac{m_r^2}{4 M_W^2} 
    + 2 \frac{m_r^2}{s} \right) \frac{m_r^2}{s - m_r^2}
\nonumber \\ & &{} \qquad\qquad
    - \left( \frac{M_W^2}{m_r^2} + 1 + \frac{m_r^2}{4 M_W^2} 
    - 2 \frac{m_r^2}{s} \right) \frac{m_r^2}{s} 
        \ln \left( 1 + \frac{s}{m_r^2} \right) \Bigg] \; .
\label{a0-eq}
\end{eqnarray}
The leading order term for $s \gg M_W^2,m_r^2$ can be rewritten
compactly as
\begin{eqnarray}
a_0|_{\rm leading} &=& - \frac{1}{192 \pi} \frac{s}{\Lambda^2} \; .
\end{eqnarray}
Thus, the radion mass does not regularize the 
bad high energy behavior of the partial wave amplitude
because there are no particular symmetry relations between
the radion and the Goldstone bosons.  However, we have already 
argued that our 4-d effective theory is valid for energies 
only up to about the cutoff scale $\Lambda$. Under the condition 
$s,m_r^2 < \Lambda^2$, the radion contributions will not
saturate unitarity and thus no significant bounds can be 
obtained.

\subsection{Radion interactions with electroweak Goldstone bosons}
\label{radion-gold-sec}

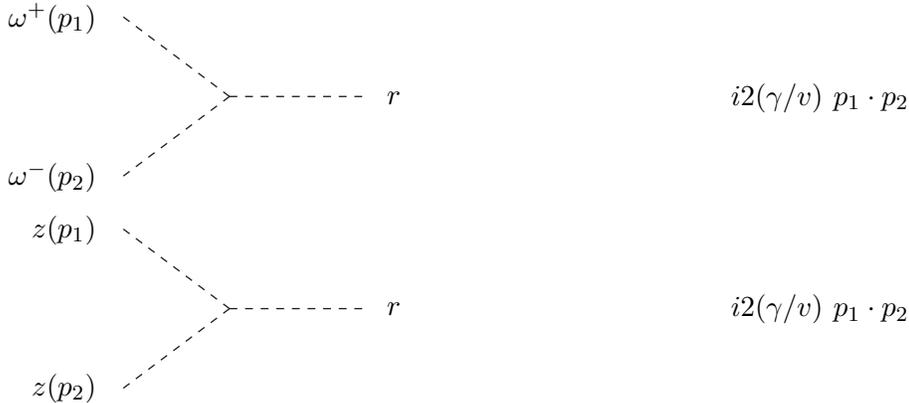
\begin{figure}[t]
\begin{picture}(440,140)
%
  \Text(      60, 140 )[r]{$\omega^+(p_1)$}
  \Text(      60,  80 )[r]{$\omega^-(p_2)$}
  \DashLine(  70, 140 )( 110, 110 ){3}
  \DashLine(  70,  80 )( 110, 110 ){3}
  \DashLine( 110, 110 )( 160, 110 ){3}
  \Text(     170, 110 )[l]{$r$}
  \Text(     300, 110 )[l]{$i 2 (\gamma/v) \,\, p_1 \cdot p_2$}
  \Text(      60, 60 )[r]{$z(p_1)$}
  \Text(      60,  0 )[r]{$z(p_2)$}
  \DashLine(  70, 60 )( 110, 30 ){3}
  \DashLine(  70,  0 )( 110, 30 ){3}
  \DashLine( 110, 30 )( 160, 30 ){3}
  \Text(     170, 30 )[l]{$r$}
  \Text(     300, 30 )[l]{$i 2 (\gamma/v) \,\, p_1 \cdot p_2$}
\end{picture}
\caption{Feynman rules for the Goldstone bosons $\omega^\pm$ and $z$
with all momenta incoming.}
\label{feyn-rule-Goldstone-fig}
\end{figure}

The above analysis utilized the the gauge boson -- radion 
Feynman rules while attaching longitudinally polarized $W$s 
to the external lines.  It is instructive to see the same 
calculation done more directly by simply considering the couplings 
of the radion to the Goldstone bosons.  This can be done by 
starting with the fundamental couplings of the radion
in Eq.~(\ref{trace-eq}), but writing out only the scalar
kinetic and potential terms.  This is equivalent to turning off
the electroweak gauge couplings, $g = g' \ra 0$.  In fact, 
only the kinetic terms are relevant
\begin{eqnarray}
S &=& \int d^4 x \, \partial_\mu H^{\dag} \partial^\mu H 
\end{eqnarray}
where $H = (-i\omega^+, \textfrac{1}{\sqrt{2}} (h + i z))$
and $\omega^\pm$, $z$ are the charged and neutral Goldstone bosons 
respectively.
This results in the following interactions between the 
Goldstone bosons and the radion at ${\cal O}(r)$,
\begin{eqnarray}
{\cal L} &=& - \frac{2 \gamma}{v} r \partial_\mu w^+ \partial^\mu w^- 
             - \frac{\gamma}{v} r \partial_\mu z \partial^\mu z \; ,
\end{eqnarray}
leading to the Feynman rules shown in Fig.~\ref{feyn-rule-Goldstone-fig}.
Using these Feynman rules we obtain the same leading order amplitude 
given by the terms in the first line of Eq.~(\ref{radion-amp-eq}).

At this point, we should comment on this equivalence.  
The physical Higgs is part of the full Higgs doublet of the SM, 
and so its interactions are enforced by the electroweak symmetry.
It is not immediately obvious that the Goldstone boson 
approximation automatically applies for radion interactions, since 
the latter are dimension five operators added to SM Lagrangian.  In fact,
naively one might add a curvature-Higgs mixing term to the scalar 
Lagrangian that is arranged to cancel the kinetic terms in the 
energy momentum tensor.  This addition ``improves'' the 
energy momentum tensor by rendering it classically conformally
invariant.  However, as we will see in the next Section, the
addition of curvature-Higgs mixing can be consistently treated
in either prescription.  Hence, the Goldstone boson approximation 
does indeed apply for the case with the radion, including when mixed 
with the Higgs.

Since the radion contributions to the gauge boson scattering amplitude 
are suppressed by $\gamma^2 \propto 1/\Lambda^2$, there is no region
where a large contribution is expected in the 4-d effective theory, 
despite the ${\cal O}(s/M_W^2)$ divergence.  This same result
can be trivially applied to all other gauge boson scattering 
amplitudes involving the exchange of an (internal) radion.  
Fundamentally this result is straightforward to see from the
radion interactions -- the radion contribution is always
suppressed by $1/\Lambda$ instead of $1/v$.
There is, however, one important difference between the Higgs and
the radion interactions:  At one-loop the radion
couples to trace anomaly gauge interaction \cite{GRW2,BaeFeb2000}.  
Here, the strength
of the interaction is $\gamma b_a g_a^2/16 \pi^2$, where $b_a$ is 
one-loop beta function coefficient for the $a$ coupling.  This is
generally much larger than the analogous one-loop coupling to the Higgs
[at least for $a=$SU(3)],
scaling roughly with the number of heavy fermions running in the loop.
Nevertheless, this is still \emph{suppressed} relative to the size 
of tree-level couplings and thus does not give rise to any new unitarity 
problems.

\section{Curvature-Higgs mixing}
\label{curv-scalar-sec}

The couplings of the radion that we have used to calculate gauge
boson scattering were independent of the Higgs couplings.  This need
not be the case, however.  In particular, on the TeV brane we ought
to add all terms to the action that are not forbidden by symmetries.  
One important term is a curvature-Higgs mixing operator \cite{GRW2}
\begin{eqnarray}
S^{\rm mixing} &=& \int d^4 x\ \sqrt{-g_{\rm ind}} \,\, \xi \, 
   H^\dagger H {\cal R}^{(4)}(g_{\rm ind}) \; ,
\end{eqnarray}
where $\xi$ is the dimensionless coefficient of this operator,
$g_{\rm ind}$ is the induced metric on the TeV brane and
${\cal R}^{(4)}(g_{\rm ind})$ is the 4-d Ricci scalar written 
explicitly as a function of the induced metric.  In theories with
a flat extra dimension, this term gives rise to a tiny $1/\Mpl$ 
suppressed mixing between the radion and the Higgs.  
In RS1, however, this mixing is suppressed by the inverse warped
Planck scale $1/\Lambda$.  The curvature-Higgs mixing operator 
induces kinetic mixing between the radion and the Higgs \cite{GRW2,CGK}
\begin{eqnarray}
S^{\rm mixing} &=& \int d^4 x \left[ 6 \xi \gamma h\ \Yfund r 
   + 3 \xi \gamma^2\ (\partial r)^2 \right] \; .
\end{eqnarray}
This mixing can be diagonalized by appropriate field redefinitions
and rotations, resulting in the following relations between the
interaction and mass eigenstates \cite{CGK}
\begin{equation}
\begin{array}{rcl}
        h & \rightarrow & A h_m + B r_m \\
-\gamma r & \rightarrow & C h_m + D r_m
\end{array} \label{mixed-eq}
\end{equation}
where
\begin{equation}
\begin{array}{rcl}
A &=& \cos \theta - \frac{6\xi \gamma }{Z} \sin \theta \\
B &=& \sin \theta + \frac{6\xi \gamma }{Z} \cos \theta \\
C &=& \frac{\sin \theta}{Z} \gamma \\
D &=& -\frac{\cos \theta}{Z} \gamma
\end{array}
\label{ABCD-eq}
\end{equation}
and 
\begin{eqnarray}
         Z^2 &=& 1 + 6 \xi \gamma^2 \left( 1 - 6 \xi \right) 
         \label{Z-eq} \\
\tan 2\theta &=& 12 \xi \gamma Z \frac{m_h^2}{m_r^2-m_h^2 
                 \left( Z^2 - 36 \xi^2 \gamma^2 \right)} \; . 
\label{tantheta-eq}
\end{eqnarray}
The mass eigenstates are written as $h_m$ and $r_m$, although we
should emphasize that these are mixed scalars with interactions 
that do not necessarily resemble the Higgs and radion interaction
eigenstates.  Nevertheless, in the limit that $\xi \rightarrow 0$, 
the mixing matrix becomes
\begin{eqnarray}
\left( \begin{array}{cc} A & B \\ C & D \end{array} \right) 
    \stackrel{\xi \rightarrow 0}{\longrightarrow}
    \left( \begin{array}{cc} 1 & 0 \\ 0 & -\gamma \end{array} \right)
\end{eqnarray}
and hence $h \rightarrow h_m$ and $r \rightarrow r_m$.

\begin{figure}[t]
\begin{picture}(440,140)
%
  \Photon(    70, 140 )( 110, 110 ){5}{5} 
  \Photon(    70,  80 )( 110, 110 ){5}{5} 
  \DashLine( 110, 110 )( 160, 110 ){3}
  \Text(      60,  80 )[r]{$V_\mu$}
  \Text(      60, 140 )[r]{$V_\nu$}
  \Text(     165, 110 )[l]{$h_m$}
  \Text(     300, 110 )[c]{$i \frac{2 M_V^2}{v} \left( A + C \right) 
                           \eta_{\mu\nu}$}
%
  \Photon(    70,  60 )( 110,  30 ){5}{5} 
  \Photon(    70,   0 )( 110,  30 ){5}{5} 
  \DashLine( 110,  30 )( 160,  30 ){3}
  \Text(      60,   0 )[r]{$V_\mu$}
  \Text(      60,  60 )[r]{$V_\nu$}
  \Text(     165,  30 )[l]{$r_m$}
  \Text(     300,  30 )[c]{$i \frac{2 M_V^2}{v} \left( B + D \right) 
                           \eta_{\mu\nu}$}
\end{picture}
\caption{Feynman rules for three-point gauge boson couplings
including curvature-Higgs mixing.}
\label{mixed-feyn-rules-fig}
\end{figure}
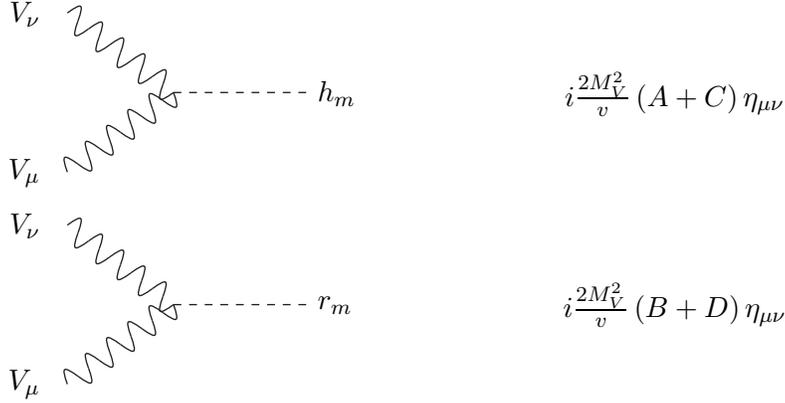

The Feynman rules for the mixed states can be easily read off from
Eq.~(\ref{rww-eq}) after substituting Eq.~(\ref{mixed-eq}).  For 
completeness, we show them in Fig.~\ref{mixed-feyn-rules-fig}.

\subsection{Mass independent contributions to $a_0$}
\label{radioneffects-curv-sec}

Since the Higgs interaction is modified, the cancellation between
the Higgs contribution shown in Fig.~\ref{W-scat-feyn-diag-fig}(d),(e) 
and the gauge contribution Fig.~\ref{W-scat-feyn-diag-fig}(a),(b),(c) 
is no longer complete.  There are now several important contributions:
the two mixed scalar (Higgs and radion) contributions, and the incompletely
canceled gauge boson piece.  The leading order, mass independent 
amplitude for the sum of these contributions is
\begin{eqnarray}
-i {\cal M} &=& - g^2 \left( (A + C)^2 + (B + D)^2 - 1 \right) 
   \left[ \frac{s}{4 M_W^2} + \frac{t}{4 M_W^2} + 2 \frac{t}{s} \right] \; .
\label{M-mi-eq}
\end{eqnarray}
Using Eqs.~(\ref{ABCD-eq})--(\ref{tantheta-eq}), the coefficient 
can be written as 
\begin{eqnarray}
(A + C)^2 + (B + D)^2 &=& 1 + \left( \frac{1 - 6 \xi}{Z} \gamma \right)^2 
\label{ACBD-relation-eq}
\end{eqnarray}
and so Eq.~(\ref{M-mi-eq}) reduces to
\begin{eqnarray}
-i {\cal M} &=& - g^2 \left( \frac{1 - 6 \xi}{Z} \gamma \right)^2
   \left[ \frac{s}{4 M_W^2} + \frac{t}{4 M_W^2} + 2 \frac{t}{s} \right] \; .
\end{eqnarray}
Again, the same result can be obtained through the Goldstone
boson approximation, as shown in the Appendix.

\begin{figure}[t]
\centering
\hspace*{0in}
\epsfxsize=5.0in
\epsffile{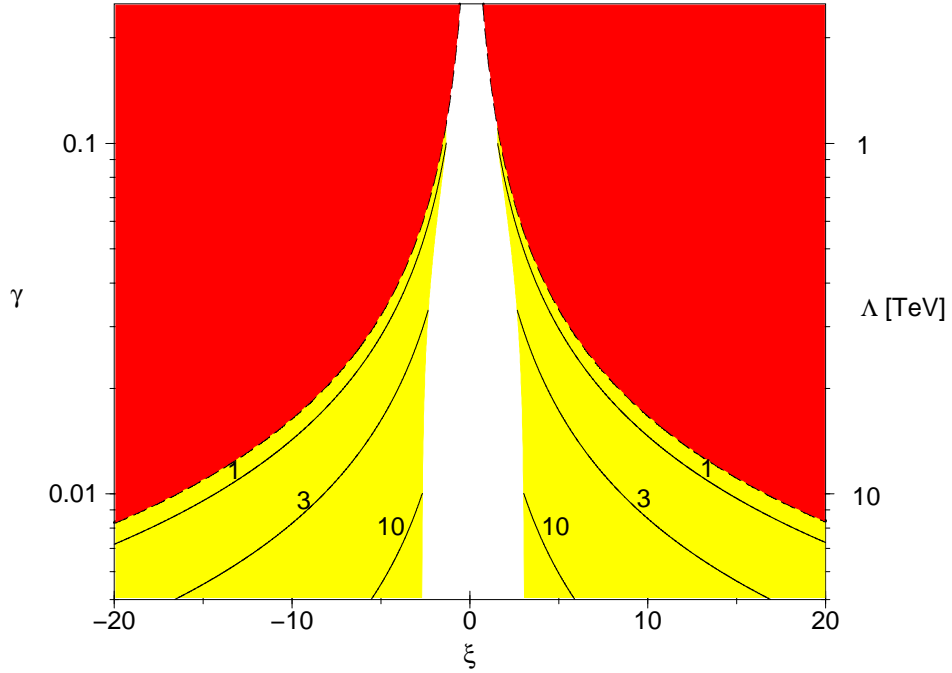}
\caption{Bound on $\gamma = 1/(\sqrt{6} \Lambda)$ as a function of $\xi$.
The dark shaded region (red) is excluded by requiring that
kinetic terms for the Higgs and radion are positive 
(the crossover is shown by the dashed line).
The light shaded region (yellow) is excluded by requiring
that the gauge boson partial wave scattering amplitude 
does not exceed the perturbative unitarity bound prior to the
TeV brane cutoff scale.  The scattering energy $\sqrt{s}$ in TeV 
at which perturbative unitarity is violated is indicated 
by the contours in this region.} 
\label{bound-fig}
\end{figure}
For small $\xi$ and $\gamma$, the coefficient of the $s/M_W^2$ term becomes
\begin{eqnarray*}
(1 - 12 \xi) \gamma^2
\end{eqnarray*}
and so there is no significant enhancement of the scalar contribution
to the gauge boson scattering amplitude.  However, if $\xi$ is 
proportional to $1/\gamma$, then the coefficient may not be suppressed 
by $\gamma^2$.  For instance, suppose $\xi = -\epsilon/6 \gamma$ with 
$\epsilon \lsim 1$, then the coefficient becomes
\begin{eqnarray*}
\frac{\epsilon^2}{1 - \epsilon^2 + \epsilon \gamma}
\end{eqnarray*}
which is ${\cal O}(1)$ and \emph{not} suppressed by $\gamma^2$.
Hence, for a large curvature-Higgs mixing coefficient 
there can be large contributions to the gauge boson scattering 
amplitude.  The partial wave amplitude 
for the mass independent (``mi'') terms is 
\begin{eqnarray}
a_0^{\rm mi} &=& 
- \frac{g^2}{16 \pi} \left( \frac{1 - 6 \xi}{Z} \gamma \right)^2
\left[ \frac{s}{8 M_W^2} - \frac{3}{2} \right] \; .
\label{a0-mi-eq}
\end{eqnarray}
Requiring that this contribution to the partial wave amplitude does
not exceed the perturbative unitarity bound implies an upper bound
on the gauge boson scattering energy, for moderate or large $\xi$.
We show the bound as a function of $\xi$ in Fig.~\ref{bound-fig}.
This is one of the central results of this paper.
If we require that our 4-d effective theory is perturbative
up to the TeV brane cutoff scale, meaning
\begin{eqnarray}
a_0(s) &<& \frac{1}{2} \qquad 
    \mbox{for all} \; s \; \mbox{up to} \; \Lambda^2 \; ,
\end{eqnarray}
then the light shaded region (yellow) is excluded.  We find that 
$|\xi|$ must be less than about $2.7$, for perturbative unitarity to 
be respected in our 4-d effective theory independent of the cutoff scale.   
As expected, for small $|\xi|$, the mass independent contribution 
never exceeds the perturbative unitarity bound, which is consistent 
with the discussion in Sec.~\ref{W-scat-sec}.  For $|\xi| \gsim 2.7$,
we can identify the gauge boson scattering energy at which perturbation 
theory breaks down, which is shown by the contours.
The contours halt at small $|\xi|$ when $\sqrt{s} = \Lambda$.
The dark shaded region is excluded by requiring that the
kinetic terms for the radion and the Higgs are positive definite,
and so $Z^2 > 0$ in Eq.~(\ref{Z-eq}) \cite{CGK}.
No significant bound is obtained for the region $\Lambda \lsim 1$ TeV
($\gamma \gsim 0.1$), since at best the leading order term is 
${\cal O}(1 \; {\rm TeV}^2/M_W^2)$, and we already know
that the ${\cal O}(m_h^2/M_W^2)$ term with $m_h \sim 1$ TeV is 
marginally allowed in the SM.

\begin{figure}[!t]
\hspace*{0in}
\centerline{
\epsfxsize=0.55\textwidth
\epsffile{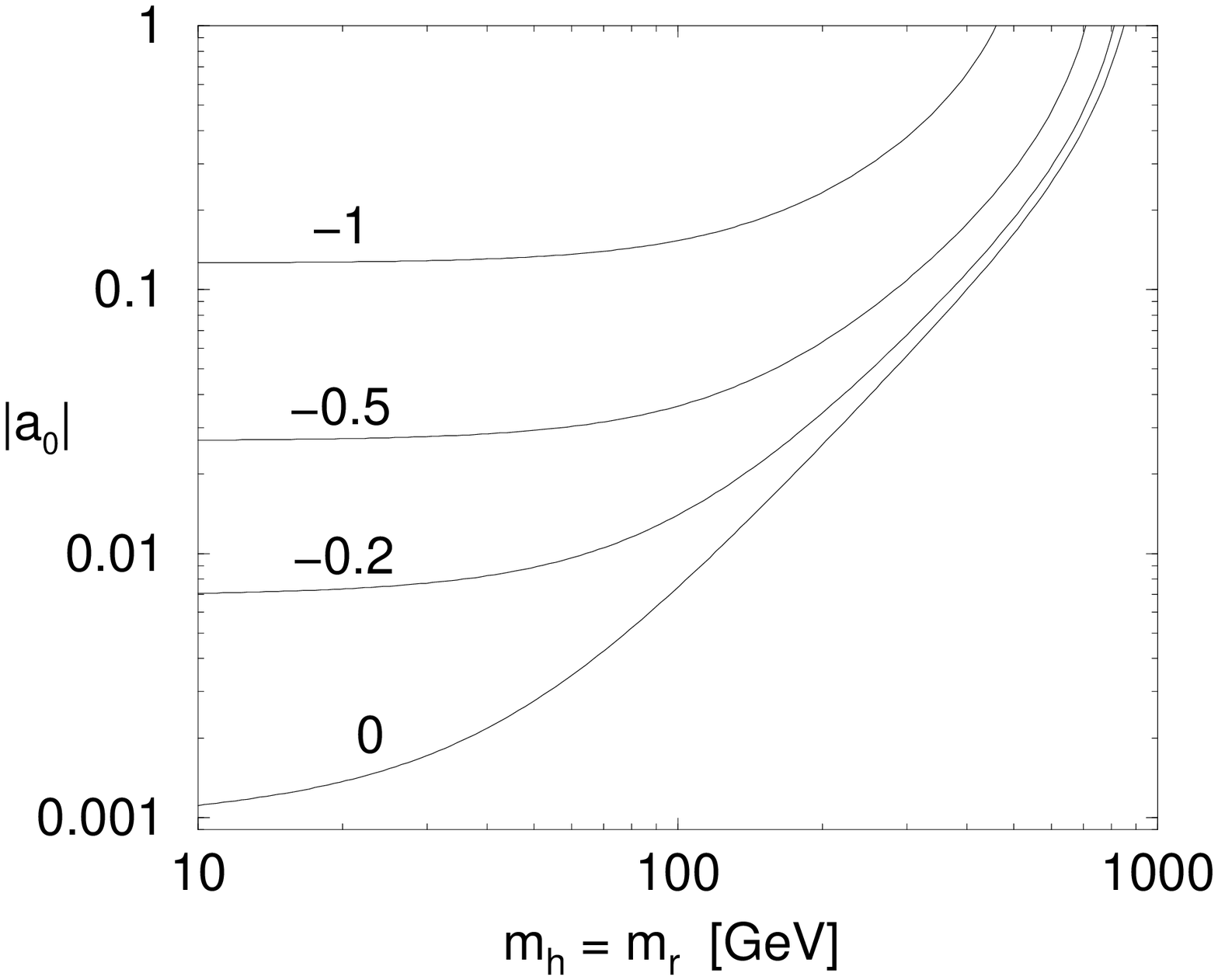}
\hfill
\epsfxsize=0.55\textwidth
\epsffile{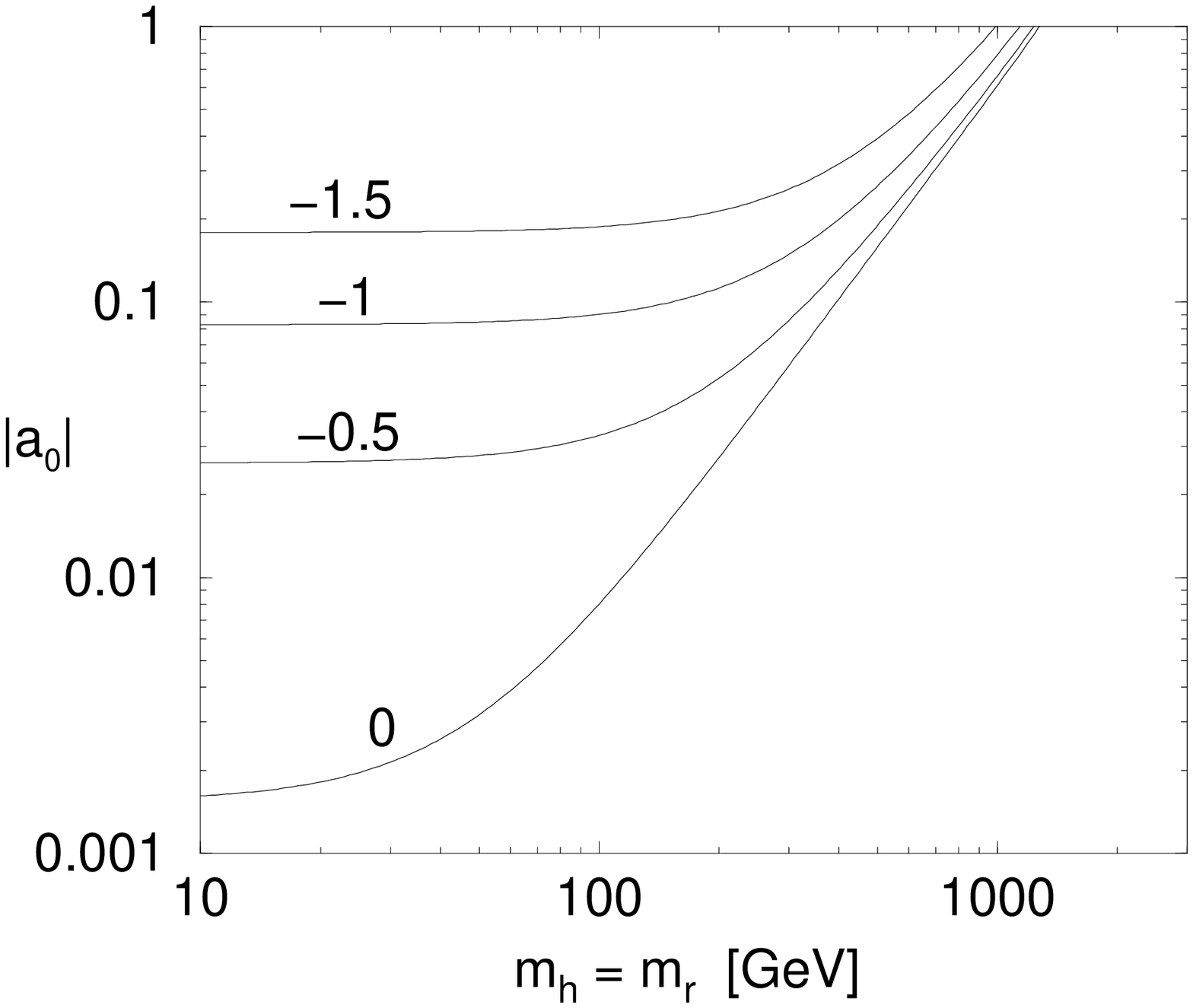}
}
\caption{The contribution to the $j=0$ partial wave amplitude $a_0$
from all tree-level diagrams, including the mass effects of Higgs 
and the radion.  The $x$-axis corresponds to the interaction eigenstate 
masses of the Higgs and radion (taken to be equal for illustration).
The scattering amplitude was evaluated at the maximal energy 
$\sqrt{s} = \Lambda = (1,3)$ TeV for the graph on the (left,right).
Contours of various values of the curvature-Higgs mixing parameter 
$\xi$ are shown.}
\label{mass-effect-fig}
\end{figure}

\subsection{Mass dependent contributions to $a_0$}

Up to now we have considered the mass independent terms to
gauge boson scattering.  The full result, including mass effects, is 
\begin{eqnarray}
a_0^{\rm total} &=& a_0^{\rm mi}
    - (B + D)^2 \Bigg[ \frac{m_{r_m}^2}{2 M_W^2}
    \left( \frac{M_W^2}{m_{r_m}^2} + 1 + \frac{m_{r_m}^2}{4 M_W^2} 
    + 2 \frac{m_{r_m}^2}{s} \right) \frac{m_{r_m}^2}{s - m_{r_m}^2}
\nonumber \\ & &{} \qquad\qquad\qquad\;\;\;
    - \left( \frac{M_W^2}{m_{r_m}^2} + 1 + \frac{m_{r_m}^2}{4 M_W^2} 
    - 2 \frac{m_{r_m}^2}{s} \right) \frac{m_{r_m}^2}{s} 
        \ln \left( 1 + \frac{s}{m_{r_m}^2} \right) \Bigg]
\nonumber \\ & &{} 
    - (A + C)^2 \Bigg[ \frac{m_{h_m}^2}{2 M_W^2}
    \left( \frac{M_W^2}{m_{h_m}^2} + 1 + \frac{m_{h_m}^2}{4 M_W^2} 
    + 2 \frac{m_{h_m}^2}{s} \right) \frac{m_{h_m}^2}{s - m_{h_m}^2}
\nonumber \\ & &{} \qquad\qquad\quad\;
    - \left( \frac{M_W^2}{m_{h_m}^2} + 1 + \frac{m_{h_m}^2}{4 M_W^2} 
    - 2 \frac{m_{h_m}^2}{s} \right) \frac{m_{h_m}^2}{s} 
        \ln \left( 1 + \frac{s}{m_{h_m}^2} \right) \Bigg]
\label{a0-total-eq}
\end{eqnarray}
where $m_{h_m},m_{r_m}$ are the physical masses of the mixed
Higgs/radion scalars.  Notice that the scalar mixing coefficients 
are positive definite, and in fact from Eq.~(\ref{ACBD-relation-eq})
it is obvious that $(A + C)^2 > \textfrac{1}{2}$ or 
$(B + D)^2 > \textfrac{1}{2}$, and so there are always significant
contributions to the partial wave amplitude from these mass terms.
We illustrate this in Fig.~\ref{mass-effect-fig} 
in two cases with $\sqrt{s}=\Lambda=1, 3$ TeV
by varying the masses and the curvature-Higgs mixing $\xi$, while
fixing the interaction eigenstate
masses to be equal.
One has to take care in doing the calculation, since now there
are poles in $1/(s - m^2)$ for $m=m_{h_m}$ and $m=m_{r_m}$.  With
maximal $s = \Lambda^2$, the effects of these poles show up only 
when the mass is ${\cal O}(\Lambda)$, which is
already in the region in which we cannot reliably calculate.
The result is that the ``allowed'' region shown in Fig.~\ref{bound-fig} 
is further reduced by including the finite 
mass effects of radion and the Higgs.

\begin{figure}[t]
\centering
\hspace*{0in}
\epsfxsize=4.0in
\epsffile{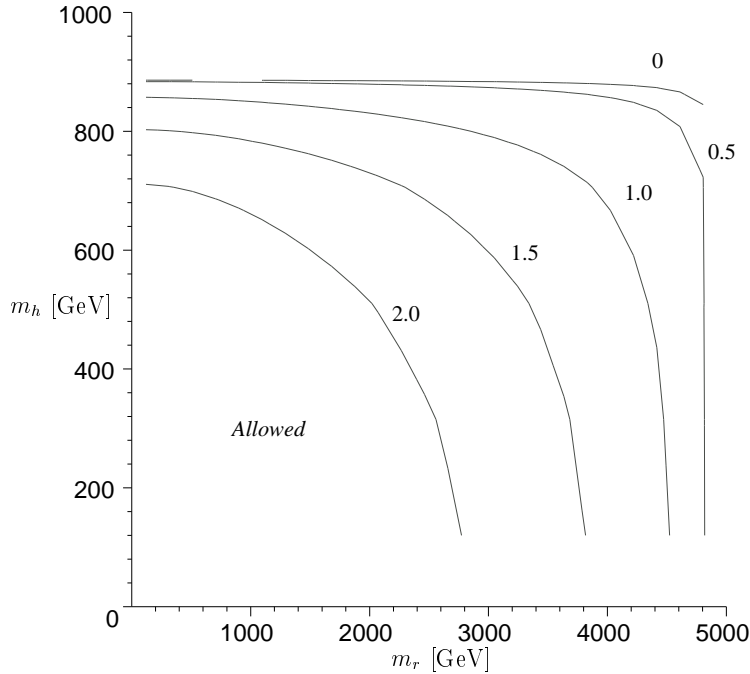}
\caption{Constraints on the radion and Higgs masses from
the perturbative unitarity bound.  For this graph, we set 
$\sqrt{s} = \Lambda = 5$ TeV\@.  Contours of various values of 
$\xi$ are shown with $a_0 = 0.5$.  The allowed region lies below
and to the left of the contours.}
\label{mass-bound-fig}
\end{figure}

The above results were obtained by assuming the masses of the
two scalars are (approximately) equal.  Even when the masses
are widely separated, we find that the unitarity bound is 
reduced.  We illustrate this in Fig.~\ref{mass-bound-fig}
where we show contours of $|a_0| = 0.5$ as a function of 
the Higgs and radion masses, for the choice $\sqrt{s} = \Lambda = 5$ TeV
and various values of $\xi$.   
The interaction eigenstate masses are shown on the axes, although they are
nearly the same as the the physical masses throughout the plot.
Thus, requiring perturbative unitarity is not violated places
upper bounds on the masses of the mixed scalars (radion and Higgs),
in regions of ($\gamma$,$\xi$) parameter space \emph{allowed} 
by Fig.~\ref{bound-fig}.

\begin{figure}[t]
\centering
\hspace*{0in}
\epsfxsize=4.0in
\epsffile{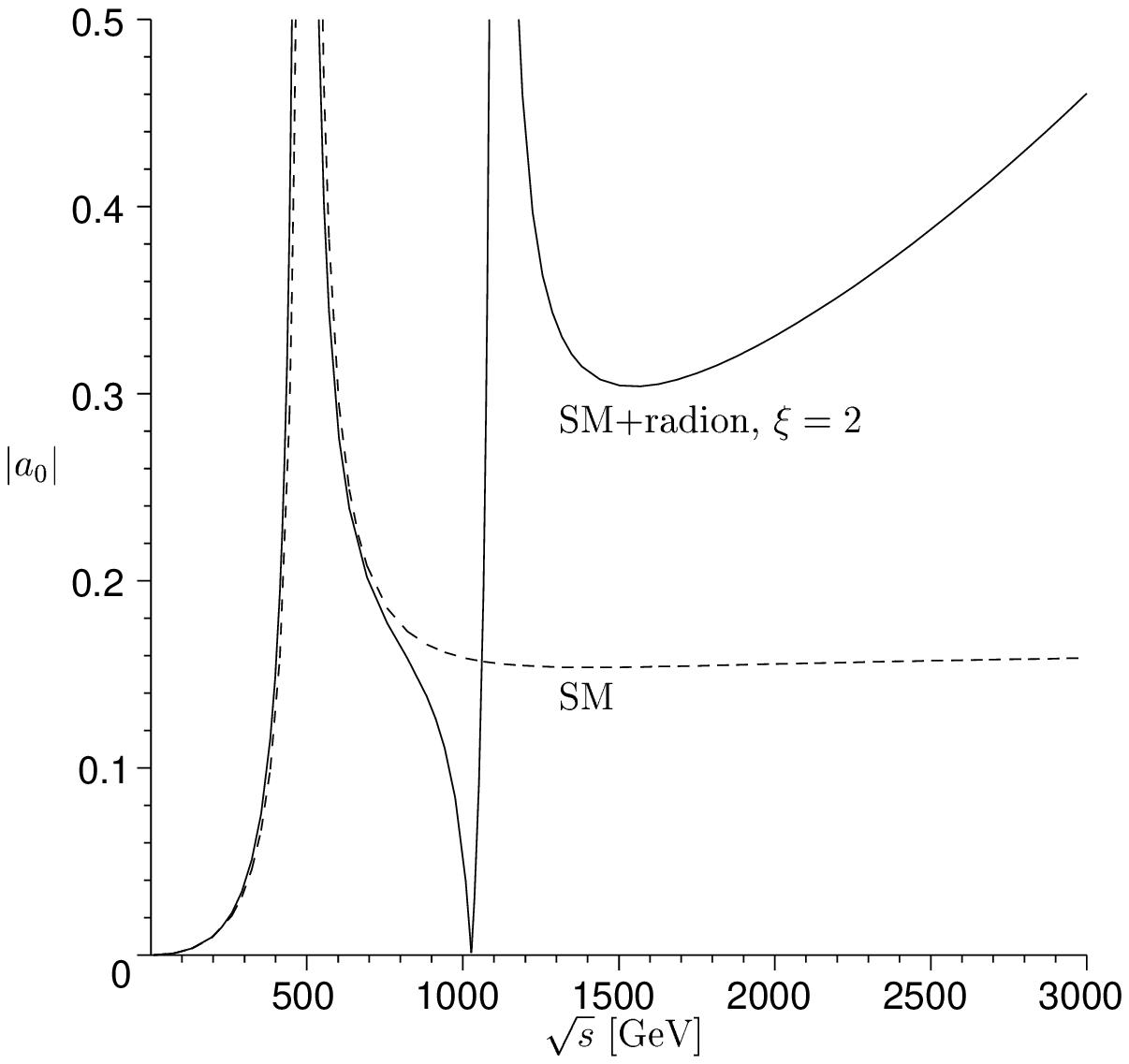}
\caption{The partial wave amplitude as a function of $\sqrt{s}$.
The dashed line is the SM for $m_h = 500$ GeV (width effects
were \emph{not} taken into account).  The solid line shows
the behavior for the same Higgs mass but with a radion of mass
$\sim 2 m_h$ with curvature-Higgs mixing in the spectrum.  
Notice that destructive interference occurs between 
$m_h < \sqrt{s} < m_r$, due to the differing sign
of the (leading) pole contributions, which contrasts with the
asymptotically flat behavior of the partial wave amplitude in the SM.}
\label{amp-func-s-fig}
\end{figure}
The contributions to the partial wave amplitude above the Higgs and 
radion mass poles always constructively interfere, and so the
presence of a radion universally increases the size of the partial
wave amplitude.  Below the mass of the Higgs and the radion
$\sqrt{s} \ll m_h, m_r$,
the partial wave amplitude is given by Eq.~(\ref{part-eq})
with no dependence on the masses of the scalars, as the 
decoupling theorem requires.  The remaining region, namely
$m_h < \sqrt{s} < m_r$ (or $m_r < \sqrt{s} < m_h$) has the
interesting behavior that the contribution from the Higgs
and the radion have opposite sign due to the differing sign
of the pole $m^2/(s - m^2)$.  While this does not give
rise to any new unitarity problems nor does it relax the above
constraints, it is an interesting \emph{distinction} as compared
with the high energy behavior of the partial wave amplitude of the Higgs.
In Fig.~\ref{amp-func-s-fig} we illustrate the differing behavior 
of the partial wave amplitude as a function of scattering energy.  
Notice that for scattering energies lower than the radion mass,
the partial wave amplitude drops rapidly.  This should be compared
against the SM (or the ``SM + radion'' no mixing case) which
asymptotes to a constant at large energy.  

\section{Conclusions}
\label{conclusions-sec}

We have calculated the tree-level contribution to gauge boson 
scattering including the radion in the 4-d effective theory,
valid below the cutoff scale $\Lambda = e^{-kL} \Mpl$,
obtained from the Randall-Sundrum solution to the hierarchy problem.
In general, the radion provides a new contribution that is linearly 
divergent in $s$, going as $s/\Lambda^2$.  This was calculated
using longitudinally polarized gauge bosons with the three-point
$W_\mu^+ W_\nu^- r$ couplings, and independently using the eaten
Goldstone bosons with their $\omega^+\omega^- r$ couplings.
However, since our effective
theory is only valid up to of order $\Lambda$, no significant
bounds can be obtained when the Higgs and radion are mass
eigenstates with no mixing.  This linear contribution is exactly
analogous to the Higgs, but unlike the Higgs there is no gauge 
cancellation between this term and ordinary gauge boson exchange.

Including curvature-Higgs mixing into the 4-d effective theory
dramatically affects the gauge boson scattering cross section.  
In particular, we found that the mixing coefficient $|\xi|$ 
must be less than $3$ so that perturbative unitarity 
is not violated \emph{prior} to reaching the TeV brane cutoff 
scale $\Lambda$, independent of the mass of the radion and Higgs
as seen in Fig.~\ref{bound-fig}.  It is interesting to 
remark that with the curvature-Higgs mixing coefficient 
$\xi \sim -1/(6 \gamma)$, electroweak precision constraints can
be satisfied with $m_h,m_r \gg M_W$ \cite{CGK}.  We find, however,
that this large mixing implies the theory must also become 
non-perturbative at a scale significantly \emph{below} the TeV brane 
cutoff scale.
We also calculated the partial wave 
amplitude including the radion and Higgs masses, and we found 
that the allowed region of $(\gamma,\xi)$ satisfying perturbative
unitarity is further reduced.  Mass bounds on the Higgs boson and
the radion can be inferred for large $\xi$, 
as shown in Fig.~\ref{mass-bound-fig}.
Via our explicit calculations, we verified the validity of
the Goldstone boson equivalence theorem with the existence
of the curvature-Higgs mixing.

\section*{Acknowledgments}
\indent

G.D.K.\ thanks the theoretical physics group at LBL where part of
this work was completed.
This work was supported in part by the U.S. Department of Energy 
under grant number DE-FG02-95ER40896, and in part by the Wisconsin
Alumni Research Foundation.

\begin{appendix}

\section*{Appendix:  Goldstone boson calculation including mixing}

Given the action
\begin{eqnarray}
S &=& \int d^4 x \sqrt{-g} \left( g^{\mu\nu} D_\mu H^\dag D_\nu H  
   + \xi H^\dag H {\cal R}^{(4)} \right)
\end{eqnarray}
with the Higgs doublet $H = [ -i \omega^+ , \textfrac{1}{\sqrt{2}} (h + iz)]$
expressed in terms of the Goldstone bosons $\omega^\pm$ and $z$.
The induced metric expressed explicitly in terms of the radion
field is \cite{CGK}
\begin{eqnarray}
g_{\mu\nu} = e^{-2 k L - 2 (\gamma/v) r} \eta_{\mu\nu} \; .
\end{eqnarray}
It is straightforward to carry out the rescalings of the fields
$H \ra e^{-kL} H$, etc., so that the curvature-Higgs mixing term
$H^\dag H {\cal R}^{(4)}$ can be expressed as
$(\omega^+ \omega^- + \textfrac{1}{2} z^2) \Yfund r$ to ${\cal O}(r)$.  
Suitable manipulations of this term result in the Goldstone boson/radion 
interaction terms
\begin{eqnarray}
{\cal L} &=& - 2 \frac{\gamma}{v} \left( 1 - 6 \xi \right) r 
\left[ \partial_\mu \omega^+ \partial^\mu \omega^- 
       + \frac{1}{2} \partial_\mu z \partial^\mu z \right] \; .
\end{eqnarray}
We have implicitly used the Landau gauge, setting 
$\Yfund \omega^\pm = \Yfund z = 0$.  

Following the discussion in Sec.~\ref{curv-scalar-sec}, we then 
rewrite the Lagrangian in terms of the physical eigenstates.  
To see that we obtain the
same leading order mass-independent amplitude, we can make the 
simplification $m_h = m_r = 0$ in Eq.~(\ref{tantheta-eq}), 
and then from Eq.~(\ref{ABCD-eq}) it is clear that $r = r_m/Z$.  
Using the interaction Lagrangian above, with the physical radion
mass eigenstate $r_m$, the Goldstone boson partial wave amplitude 
is \emph{exactly} the same as Eq.~(\ref{a0-mi-eq}).
It is also not hard to verify the leading order mass dependent
terms given in Eq.~(\ref{a0-total-eq}) can be similarly obtained.

\end{appendix}


\end{document}